\documentclass[12pt,preprint]{aastex}
\usepackage{emulateapj5,apjfonts,onecolfloat5}

\def\sgr {SGR\,1806$-$20\,}
\def\sgrof {SGR\,0526$-$66\,}
\def\sgroo {SGR\,1900$+$14\,}
\newcommand{\CXO}{{\it Chandra}\,}

\newcommand{\XMM}{{\it XMM--Newton}\,}
 
\newcommand{\RXTE}{{\it R}XTE\,}

\def\ltsima{$\; \buildrel < \over \sim \;$}
\def\lsim{\lower.5ex\hbox{\ltsima}}
\def\loe{\lower.5ex\hbox{\ltsima}}
\def\gtsima{$\; \buildrel > \over \sim \;$}
\def\gsim{\lower.5ex\hbox{\gtsima}}
\def\goe{\lower.5ex\hbox{\gtsima}}

\def \rchisq {$\chi_{\nu} ^{2}$}
\def\ergs {erg\,s$^{-1}$}
\def\ergscm2 {erg\,s$^{-1}$cm$^{-2}$}

\shortauthors{REA ET AL.}
\shorttitle{ \CXO\, POST-FLARE VIEW OF \sgr}

\begin{document}

\twocolumn[

\doublespace

\title{A first look with \CXO\, at \sgr\, after the Giant Flare:\\
significant spectral softening and rapid flux decay}

\bigskip

\author{N. Rea\altaffilmark{1,2,*}, A. Tiengo\altaffilmark{3,4}, S. Mereghetti\altaffilmark{3}, G.L. Israel\altaffilmark{2}, S. Zane\altaffilmark{5}, R. Turolla\altaffilmark{6}, L. Stella\altaffilmark{2} }

\affil{
$^{1}$SRON--National Institute for Space Research, Sorbonnelan 2, 3584 CA, Utrecht, The Netherlands \\
$^{2}$INAF--Astronomical Observatory of Rome, Via Frascati 33, 00040, Monteporzio Catone (Roma), Italy \\
$^{3}$ Istituto di Astrofisica Spaziale e Fisica Cosmica `G.Occhialini`, via Bassini 15, 20133, Milano, Italy \\
$^{4}$ Universit\`{a} degli Studi di Milano, Dipartimento di Fisica,
v. Celoria 16, I-20133 Milano, Italy \\
$^{5}$ Mullard Space Science Laboratory, University College of London, Holbury St. Mary, Dorking Surrey, RH5 6NT, UK \\
$^{6}$ University of Padua, Physics Department, via Marzolo 8, 35131, Padova, Italy \\
$^{*}$ Marie Curie Fellow to NOVA,  The Netherlands Research School in Astronomy }

\thispagestyle{empty}
\smallskip

\begin{abstract}

We report on the results of a $\sim$30 ks \CXO\, pointing of the soft
gamma-ray repeater SGR 1806--20, the first X-ray observation with high
spectral resolution performed after the 2004 December 27 giant
flare. The source was found in a bursting active phase and with a
significantly softer spectrum than that of the latest observations
before the giant flare. The observed flux in the 2--10\,keV range was
$\sim 2.2\times 10^{-11}$ ergs cm$^{-2}$ s$^{-1}$, about 20\% lower
than that measured three months before the event. This indicates that,
although its giant flare was $\approx 100$ times more intense than
those previously observed in two other soft gamma-ray repeaters, the
post flare X-ray flux decay of \sgr\, has been much faster. The pulsed
fraction was $\sim$3\%, a smaller value than that observed before the
flare. We discuss the different properties of the post-flare evolution
of \sgr\, in comparison to those of SGR 1900$+$14 and interpret the
results as a strong evidence that a magnetospheric untwisting
occurred (or is occurring) after the giant flare.

\end{abstract}


]

\smallskip

\section{INTRODUCTION}

Soft gamma-ray repeaters (SGRs) are neutron stars which emit short
($\lesssim 1$s) and energetic ($\lesssim 10^{42}$erg s$^{-1}$)
bursts of soft $\gamma$-rays. The burst repetition time can vary
from seconds to years (\citealt{go01}). During the quiescent state
(i.e. outside bursts events) these sources are detected as
persistent X-ray emitters at a luminosity of $\sim
10^{35}-10^{36}$erg s$^{-1}$. Occasionally, SGRs emit much more
energetic ``giant flares'' ($\sim 10^{44}-10^{45}$\ergs); these
are rare events reported until recently  only on two occasions
from \sgrof \ and \sgroo \ (\citealt{ma79}, \citealt{hu99}).

Several characteristics of SGRs, including their bursting activity,
are explained in the context of the "magnetar" model
(\citealt{dt92},\citealt{td95}). Magnetars are neutron stars the
emission of which is powered by the decay of an ultra-strong magnetic field
($\sim10^{14}-10^{15}$\,G). In this model the frequent short bursts
are associated to small cracks in the neutron star crust, while the
giant flares are linked to global rearrangements of the star
magnetosphere.

On 2004 December 27 \sgr \, emitted an exceptionally powerful giant
flare, with an initial hard spike lasting 0.2 s followed by a
$\sim$600\,s long pulsating tail (\citealt{bo04}, \citealt{hu04},
\citealt{ma04}). The prompt emission saturated almost all $\gamma$-ray
detectors, except for those on the GEOTAIL spacecraft which provided a
reliable measurement of the peak intensity (\citealt{to05}).  The
isotropic luminosity above 50 keV was $\sim 6.47 \times
10^{47}$\ergs\, (for a distance of 15 kpc), hundreds of times higher
than that of the two giant flares previously observed from other
SGRs. Following this event, afterglow emission, similar to that
commonly observed in gamma-ray bursts, has been observed in the radio
band with a resolved extended structure (\citealt{ca05},
\citealt{ga05}), and possibly also at hard X-ray energy
(\citealt{sa05b}). The extremely accurate localization
($\sim$0.1\arcsec) obtained with the radio data made possible the
identification of a variable infrared counterpart (\citealt{kosu05},
\citealt{isra05}).

Here we report the results of a \CXO\, Director's Discretionary
Time observation of \sgr\, which provided the first X-ray data
set with high spectral resolution after the giant flare.

\section{OBSERVATION}

\CXO\,  observed \sgr\,  for $\sim$30\,ks with the Advanced CCD
Imaging Spectrometer (ACIS) instrument on 2005 February 8. In order to
avoid pile-up and perform pulse phase resolved spectroscopy, the
source was observed in the Continuous Clocking (CC) mode, which
provides a time resolution of 2.85\,ms and imaging along a single
direction. The source was positioned in the back-illuminated ACIS-S3
CCD at the nominal target position. Standard processing of the data
was performed by the \CXO\, X-ray Center to Level 1 and Level 2
(processing software DS 7.5.0.1). The data were reprocessed using the CIAO
software (version 3.2) and the \CXO\, calibration files (CALDB version
3.0.0).

Since in the CC mode the events are tagged with the  times of the
frame store, we  corrected the times for the variable delay due to
the spacecraft dithering and telescope flexure, starting from
Level 1 data, and assuming that all photons were originally
detected at the target position\footnote{see the  {\em Chandra
Science Threads} at http://asc.harvard.edu/ciao/threads/index.html
for details}. We filtered the data to exclude events with ASCA
grades 1, 5 and 7, hot pixels, bad-columns and possible afterglow
events (residual charge from the interaction of a cosmic ray in
the CCD). In the data processing and analysis we always used the
specific bad-pixel file of this observation rather than those
provided with the standard calibration files. After such filtering
the observing time was 29.1\,ks.

\section{RESULTS}

\smallskip

\subsection{Timing}

In order to carry onto timing analysis we extracted the events in the
1--10\,keV energy range from a region of $5\times5$ pixels around the
source position and corrected their arrival times to the barycenter of
the Solar System. We looked for the presence of bursts by binning the
counts in intervals of 0.2 s and searching for excesses above a count
threshold corresponding to a chance occurrence of 0.1\% (taking into
account the total number of bins).  In this way we identified a single
burst, lasting about 0.5 s, at 00:38:25 UT of February 9.

The relatively poor statistics and small pulsed fraction did not
permit to determine the pulse period independently from the \CXO\,
data. We therefore adopted a period of P=7.560023\,s, measured during
an almost simultaneous \RXTE\, observation (P. Woods, private
communication). For this period the Z$^{2}_{m}$ test (Buccheri et
al. 1983) gave a significance of 3.5$\sigma$ for a number of harmonics
$m=3$ (or 2.9$\sigma$ for $m=2$). The resulting pulse profiles folded
in 16 phase bins are shown in Fig.\,1 for three different energy
ranges (1--10, 1--4, and 4--10 keV). The modulation is rather low and
with some evidence for a double peaked profile and, possibly, an
energy-dependent shape. By fitting with two sine functions the pulse
profile in the total energy range we obtain pulsed fractions values of
PF$_{1}=3.0\pm1.1$\% and PF$_{2}=2.6\pm1.1$\%
\footnote{we define PF as the semi-amplitude of the two sine
functions; all errors in the text are at 90\% confidence level}, for
the fundamental and for the second harmonic, respectively.


\begin{figure}[t]
\centerline{\includegraphics[height=7.5cm,width=11.0cm,angle=-90]{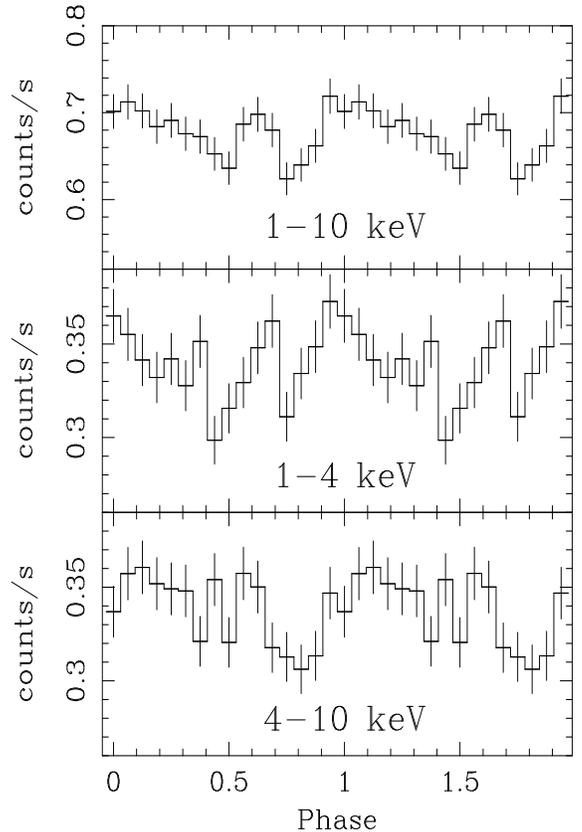}}
\caption[]{Folded pulse profiles in the total (1--10\,keV; {\em top
panel}), soft (1--4\,keV; {\em middle panel}) and hard
(4--10\,keV; {\em bottom panel}) energy ranges. } \label{fig:efold}
\end{figure}


\smallskip

\subsection{Spectroscopy}

The source spectrum was extracted from a rectangular region of
$5\times25$ pixels around the source position, and the background
independently from a source-free region in the same chip.

Since the CC mode has not yet been calibrated, standard threads for
spectral analysis are not available, and the Timed Exposure mode
response matrices (rmf) and ancillary files (arf) are generally used.
In order to extract the rmf, we first created a weighted map with a
re-binning by a factor of 8. We then used it in the {\tt mkacisrmf}
tool, with an energy grid ranging from 0.3 to 10\,keV in 5 eV
increments.  Using this rmf and the aspect histogram created with the
aspect solution for this observation ({\tt asphist}), we generated the
appropriate arf file for the source position.  Considering that only a
few counts were detected below 2 keV, due to the high interstellar
absorption, and to uncertainties in the instrument calibrations at
these low energies, we restricted our fits to the 2--8\,keV energy
range. All the fits were performed using {\em Xspec} (version 11.3).

Equivalently good results were obtained using either a power law
(photon index $\Gamma\sim1.8$) or a thermal bremsstrahlung
model ($kT_{brem}\sim 10$\,keV), while single blackbody and
neutron star atmosphere models gave unacceptable fits
(\rchisq$\sim$2.1 in both cases). The results of the acceptable
fits are summarized in Table~1, where we report for comparison
also those obtained in September 2004, before the giant flare,
with \XMM\, (\citealt{sa05}). The best fit power law spectrum was
shown in Fig.~2. The absorption derived from the \CXO\, data was
consistent with the pre-flare value. Keeping the absorption fixed
at the \XMM\, value yielded a photon index $\Gamma=1.77\pm0.05$ and
a flux of (2.2$\pm$0.2)$\times10^{-11}$ ergs cm$^{-2}$ s$^{-1}$
(2--10 keV, not corrected for absorption).

The addition of a blackbody component to the absorbed power-law was
not required, contrary to the case of  the \XMM\, spectrum of
September 2004, which had a higher statistics. Even by performing the
fit of the \CXO\, data in the wider  0.3--10  keV range, the
inclusion of an additional blackbody component does not improve
significantly the fit (F-test probability=0.012). Note however
that the presence of a blackbody with the temperature and
normalization as seen with \XMM\, is compatible with the \CXO\,
data (see Table 1).

We performed pulse phase resolved spectroscopy by extracting the
spectra for three phase intervals, corresponding to  the rise and
the decay of the broad peak, and to the  narrow peak  (see
Fig.\,1). The resulting spectral parameters were, to within 
the uncertainties,  compatible with those of the  phase-averaged
spectrum. This is not surprising considering that the pulsed
component represents only $\sim$3\% of the total emission.

We do not find any evidence for absorption or emission features in
the source averaged and phase-resolved spectra.  We derived upper
limits as a function of the line energy and width ($\sigma_E$) by
adding Gaussian lines to the continuum model. For the
phase-averaged spectrum, the 3$\sigma$ upper limit on the
equivalent width of narrow lines is $\sim$80 eV. The corresponding
values for broad lines are 110 eV and 135 eV (for $\sigma_E$=0.1
keV and $\sigma_E$=0.2 keV, respectively).

The burst identified in the \CXO\, data does not contain enough
counts for a meaningful spectral analysis.


\begin{figure*}[t]
\centerline{\includegraphics[height=12.0cm,width=6.5cm,angle=-90]{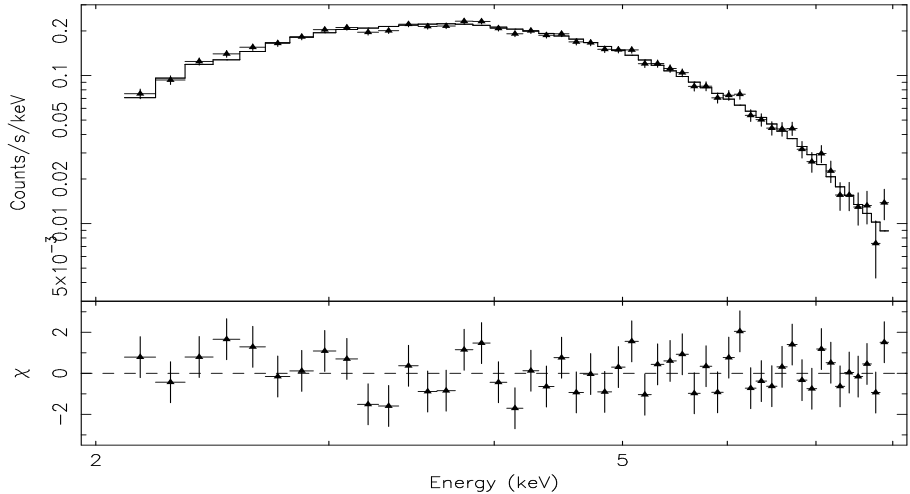}}
\caption[]{Post-flare \CXO\, \sgr\, spectrum fitted with an
absorbed power-law.} \label{fig:spec}
\end{figure*}


\smallskip

\subsection{Extended emission and structures}

In order to search for extended X-ray emission or structures around
the source, we studied the radial profiles of \sgr. Since the CC mode
has just one-dimensional imaging capability, we first generated an
image of the one-dimensional strip in the 0.3--8\,keV band, and then
subtracted an average count rate aimed to remove instrumental and
cosmic X-ray background (\citealt{mv01}).  

We then produced a one-dimensional surface brightness distribution
using the same method as for a two-dimensional radial
profile\,$^{1}$. Since calibration Point Spread Functions are not
available for the CC mode, we extracted in the same way the radial
profile from a CC-mode observation (Obs-ID 4523) of
RX\,J0806+1527. This latter source is known to be point-like and has
both an X-ray flux and spectrum rather similar to those of our target
(\citealt{gi03}).

The ratio between the radial profiles of the two sources, did not show
any evidence for a significant extended emission or structures within
30\arcsec around our target, with a 3$\sigma$ upper limit in flux of
$<10^{-14}$\ergscm2 (in the 4--30\arcsec range of radii).

\bigskip

\section{DISCUSSION}

This \CXO\ X-ray observation of \sgr\, is the first with an imaging
instrument after the December 2004 giant flare. It is therefore
interesting to compare the results with the pre-flare properties of
the source, as measured with \XMM\, in September-October 2004
(\citealt{sa05}).

The \CXO\, data clearly indicate that the spectrum softened
significantly: we obtained a power law with $\Gamma\sim$1.8. This has
to be compared with the pre-flare values $\Gamma\sim$1.2 (with the
inclusion of the blackbody) or $\Gamma\sim$1.5--1.6 (in the single
power law model, see Table~1). The flux measured with \CXO\, is
$\sim$20\% lower than the pre-flare value, but still significantly
higher than the historical flux level of $\sim$1.3$\times10^{-11}$
ergs cm$^{-2}$ s$^{-1}$ observed before the second half of
2004\footnote{Note that before the giant flare the flux was increasing
and the flux level was larger than its historical average, see
Mereghetti et al. 2005a)}. Another difference with respect to the
pre-flare properties is the smaller pulsed fraction (which changed
from about 10\% to 3\%) and the pulse profile is now double peaked.

The post-flare evolution of \sgr\, shows both similarities and
differences when compared to that of \sgroo, the only other case in
which good spectral X-ray data have been collected after a giant
flare. In particular, a significant spectral softening was observed to
accompany the post-giant flare evolution of \sgroo\, also (Woods et
al. 1999; 2001). Even though the \sgr\, giant flare was two orders of
magnitude more energetic than that of \sgroo \ (and of SGR 0526--66 as
well), it was followed by a very rapid decay of the X-ray
luminosity. We find that the source flux has dropped below the
pre-flare level after about one month, much faster than what observed
after \sgroo\, giant flare.  This suggests that the post-flare
softening, a feature common to both sources, is unrelated to the flare
energetics and the decay rate of the X-ray flux after the flare.

\sgr\, and \sgroo\, exhibit quite a different behavior also in the
evolution of their timing properties. The pulse profile of \sgroo \ 
changed from a complex, multi-peaked pattern to a simpler sinusoidal
shape. The pre-flare pulse shape has not yet been recovered, a
possible signature of a permanent rearrangement of the star
magnetosphere (Woods et al 2001).  In the case of \sgr\, the pulse
profile changed from almost sinusoidal to double-peaked.  The data
obtained during the giant flare indicate a multi-peak structure, with
a time variable and energy-dependent contribution of the different
peaks (Palmer et al. 2005, Hurley et al. 2005, Mereghetti et al.
2005b). This may indicate a different evolution of the geometry of the
magnetosphere. Moreover, while the pulsed fraction in \sgroo \ did not
change significantly after the flare, we found that in \sgr \ it
decreased by about a factor of three.

\XMM \ observations of \sgr \ carried out few months before the
giant flare have shown an increase of both the spectral hardening and
the spin-down rate with respect to historical values
(\citealt{sa05}). In the picture proposed by Thompson, Lyutikov \&
Kulkarni (2002), a twisted internal magnetic field stresses the star
solid crust, producing a progressive increase of the twist angle of
the external field lines. A giant flare is produced when the crust is
not able anymore to respond (quasi)plastically to the imparted
stresses, and finally cracks. The crustal fracturing is accompanied by
a simplification of the external magnetic field with a (partial)
untwisting of the magnetosphere. 

The spectral softening after the 2004 December 27 event appears
consistent with such a picture. In fact, the situation after the flare
is somehow opposite to what occurred before the flare, when the twist
was increasing. The sudden drop of the external twist which followed
the giant flare results in a decrease of the optical depth to resonant
cyclotron scattering in the magnetosphere and hence in a steepening of
the power-law spectrum.

The main observational consequences of a magnetospheric
untwisting, namely a decrease in the X-ray flux, a softening of the
spectrum and a decrease of the pulsed fraction (Thompson, Lyutikov \&
Kulkarni 2002; Hurley et al. 2005), appear to be present in this first
post-flare observation.

\bigskip

\acknowledgments

We thank Harvey Tananbaum for granting Director's Discretionary Time
for this observation, the whole \CXO\, team for the precious help and
the patience. We also thank P. Woods and his working group for the
information on the RXTE results. N.R. thanks M. M\'endez and L. Kuiper
for their advices, and the SRON HEA division for the
hospitality. N.R. is supported by a Marie Curie Training Grant
(HPMT-CT-2001-00245) for PhD students to NOVA, The Netherlands
Research School in Astronomy. This work was partially supported
through MIUR and ASI grants.

\onecolumn


\begin{deluxetable}{ccccccc}
\tablecolumns{7} \tablewidth{0pc} \tablecaption{Spectral results
for \sgr\,} \label{tablefits} \tablehead{ \colhead{Model} &
\colhead{$N_H$} & \colhead{$\Gamma$} & \colhead{kT} &
\colhead{$R_{bb}$\tablenotemark{a}} &
\colhead{Flux\tablenotemark{b}} &
\colhead{$\chi^2_{red}$ (d.o.f.)}  \\
\colhead{ } & \colhead{$10^{-22}$cm$^{-2}$} & \colhead{ } &
\colhead{ keV } & \colhead{km } &
\colhead{erg\,s$^{-1}$cm$^{-2}$} & \colhead{} } \startdata

\multicolumn{7} {c} {\XMM\,  Pre-flare  (Mereghetti et al. 2005a; Obs.~C) } \\
\hline
Power law    & 6.69$\pm$0.13 &  1.51$\pm$0.03&  -- & -- & 2.65 &  1.37 (72)  \\
Power law + blackbody    & 6.51\,$_{-0.27}^{+0.37}$   & 1.21\,$_{-0.12}^{+0.14}$ & 0.79\,$_{-0.12}^{+0.09}$ & 1.9\,$_{-0.3}^{+0.7}$ & 2.65  &  0.93 (70) \\
& & & &  &   & \\

\hline
\multicolumn{7} {c} {  \CXO\, Post-flare (this work) } \\
\hline
Power law  &  7.1$\pm$0.4 & 1.8$\pm$0.1 & -- & -- & 2.2  & 0.76  (129)   \\
Power law  &  6.69 fixed   & 1.77$\pm$0.05 & --  & -- &  2.2 & 0.76 (130) \\
Thermal Bremsstrahlung    &  $6.4\pm0.4$  & -- & 10.7$^{+3.6}_{-2.3}$  & -- &  2.2 & 0.78
(129) \\
Power law + blackbody     &  7.5$^{+2.3}_{-2.1}$ & 1.78$^{+0.29}_{-1.15}$  & $<$0.93   & $<$47.1 &  2.2 &  0.77 (127) \\
Power law + blackbody  & 6.51 fixed   &  1.46$\pm0.06$   &  0.79 fixed  & 1.9 fixed &  2.2  & 0.78 (130) \\
& & & &  &   & \\

\enddata
\tablenotetext{a}{For a distance of 15\,kpc; errors in the table are given at 90\% confidence level.}
 \tablenotetext{b}{In the 2--10\ keV energy band and in units of $10^{-11}$; not corrected for the absorption.}
\end{deluxetable}


\end{document}